\documentclass[nofootinbib,aps,amsmath,amssymb,a4paper,onecolumn,prd,11pt]{revtex4-1}

\usepackage[american]{babel}
\usepackage[utf8]{inputenc}
\usepackage[T1]{fontenc}

\usepackage{graphicx}
\usepackage{color}

\usepackage{units}

\usepackage[caption=false]{subfig}

\usepackage{hyperref}
\hypersetup{
   pdfnewwindow=true,     
   colorlinks=true,       
   linkcolor=black,       
   citecolor=blue,        
   filecolor=black,       
   urlcolor=blue          
}

\begin{document}

\title{\Large {\bf{Scalar Dark Matter: Direct vs.\ Indirect Detection}}}
\author{Michael \surname{Duerr}}
\author{Pavel \surname{Fileviez P\'erez}}
\author{Juri \surname{Smirnov}}
\affiliation{\\ \small{
Particle and Astro-Particle Physics Division \\
Max-Planck-Institut f\"ur Kernphysik \\
Saupfercheckweg 1, 69117 Heidelberg, Germany}
}
\begin{abstract}
We revisit the simplest model for dark matter. In this context the dark matter candidate is a real scalar field which interacts with the Standard Model particles through the Higgs portal.
We discuss the relic density constraints as well as the predictions for direct and indirect detection. The final state radiation processes are investigated 
in order to understand the visibility of the gamma lines from dark matter annihilation. We find two regions where one could observe the gamma lines at gamma-ray telescopes. 
We point out that the region where the dark matter mass is between 92 and $\unit[300]{GeV}$ can be tested in the near future at direct and indirect detection experiments.
\end{abstract}

\maketitle


\tableofcontents

\newpage

\section{Introduction}
The possibility to describe the properties of the dark matter~(DM) in the Universe with a particle candidate is very appealing. 
This idea has  motivated the theory community to propose a vast number of dark matter candidates and today we have many experiments searching for these candidates. 
The traditional way to look for dark matter is through direct detection where one expects to see the recoil energy from the scattering between the dark matter 
candidate and nucleons~\cite{Goodman:1984dc}, or from the scattering between the electrons and the dark matter. One also could see exotic signatures at the 
Large Hadron Collider~(LHC) associated with missing energy due to the production of dark matter. However, since one cannot probe the dark matter lifetime 
at colliders, this latter possibility is perhaps not the most appealing one. See Refs.~\cite{Jungman:1995df,Bergstrom:2000pn,Feng:2010gw,Abdallah:2015ter} for reviews on dark matter candidates and corresponding experimental searches. 

The annihilation of the dark matter in the galaxy into gamma rays can provide a very striking signal which can be used to determine the dark matter mass 
and  understand the dark matter distribution in the galaxy. One expects more photons in the center of the galaxy and the dark matter profile dictates 
how many photons one could expect in other regions of the galaxy for a given value of the annihilation cross section. Since the dark matter candidate does not have electric charge, the dark matter annihilation 
into monochromatic photons occurs at loop level, and it could be very difficult to observe these lines due to the continuous spectrum. See Ref.~\cite{Bringmann:2012ez} for a recent review on dark matter annihilation into gamma rays.

In the simplest dark matter model one has only a real scalar field~\cite{Silveira:1985rk}, which is stable due to the existence of a discrete symmetry. This model has only two parameters (relevant for the DM phenomenology) and one can have clear  predictions for direct and indirect detection experiments.  Since this is the minimal theory for dark matter one should investigate all the predictions to understand 
how to test this model in the near future. This model has been investigated by many groups~\cite{McDonald:1993ex,Burgess:2000yq,O'Connell:2006wi,Barger:2007im,Yaguna:2008hd,He:2008qm,Farina:2009ez,Guo:2010hq,Profumo:2010kp,Kadastik:2011aa,Djouadi:2012zc,Cheung:2012xb,Cline:2013gha,Khan:2014kba,Feng:2014vea,Kahlhoefer:2015jma,Duerr:2015mva,Han:2015hda}. 
However, only recently it has been pointed out~\cite{Duerr:2015mva} that one can observe the gamma lines from dark matter annihilation in this context due to the fact that the final state radiation~(FSR) processes are suppressed in some regions of the parameter space. This is by far not generic given a dark matter model, as photons from tree-level processes tend to dominate the spectrum.

In this article we revisit the singlet dark matter model investigating all current constraints from relic density, invisible Higgs decays, direct and indirect detection.
Here we complete the study presented in Ref.~\cite{Duerr:2015mva}. Our main aim in this article is to present the scenarios that can be tested using direct and indirect detection experiments. 
We focus on the discussion of gamma-ray lines and point out that two regions in agreement with all current experimental constraints exist where the final state radiation processes are suppressed with respect to the annihilation into photons. Thus, there is hope that actually a line could be seen over the continuum background in these regions and more information about the dark matter particle could be extracted. We do not discuss the reach of proposed future experimental searches in detail but rather focus on general features necessary to be able to distinguish lines from the continuum.

This article is organized as follows. In Section~\ref{sec:ScalarSingletDarkMatterModel} we discuss the main properties of the simplest spinless dark matter model as well as all relevant experimental constraints including the relic density.
In Section~\ref{sec:DMGammaLines} we discuss in great detail the possible gamma lines in this model and the correlation between the gamma rays coming from final state radiation and the annihilation into $\gamma \gamma$ and $Z\gamma$.
In Section~\ref{sec:Summary} we summarize our main results. In the appendix we list all relevant formulas used in this article.

\section{Scalar Singlet Dark Matter}
\label{sec:ScalarSingletDarkMatterModel}
\subsection{The Model}

In the scalar singlet dark matter model~(SDM) the dark matter candidate is a real singlet scalar field $S$ which interacts with the Standard Model~(SM) particles through the Higgs portal~\cite{Silveira:1985rk}.
The Lagrangian of this model is very simple and is given by 
 \begin{equation}
 \mathcal{L}_\text{SDM}=\mathcal{L}_\text{SM} + \frac{1}{2} \partial_\mu S \partial^\mu S -  \frac{1}{2} m_S^2 S^2 - \lambda_S S^4 - \lambda_p H^\dagger H S^2,
 \end{equation} 
where $H\sim (\mathbf{1},\mathbf{2},1/2)$ is the SM Higgs boson and $\mathcal{L}_\text{SM}$ is the usual SM Lagrangian. Once the SM Higgs acquires a vacuum expectation value, $\langle H \rangle=v_0/\sqrt{2}$ where $v_0 = \unit[246.2]{GeV}$, the physical mass of the dark matter candidate reads as
\begin{equation}
M_S^2=m_S^2+ \lambda_p v_0^2.
\end{equation}
As is usually done, we assume a discrete ${\cal Z}_2$ symmetry to guarantee dark matter stability.
Under this symmetry $S \to - S$ such that all odd terms in the scalar potential are forbidden. 

Once the electroweak symmetry is broken, the dark matter candidate $S$ can annihilate into all Standard Model particles through the portal coupling $\lambda_p$.
In this model one has only two relevant parameters for the dark matter study, the physical dark matter mass $M_S$ and the Higgs portal coupling $\lambda_p$. This is the reason why one can make definite predictions in this model once the relic density constraints are used. This model can be considered as a toy model for dark matter, but also is the perfect scenario to understand the possible predictions for different experiments and their interplay.  

\subsection{Experimental Constraints}
\subsubsection{Higgs Decays}
The most conservative, model-independent limit on the Higgs invisible decay branching ratio is set by CMS to be $\text{BR}(h \rightarrow\text{inv})<0.58$ \cite{Chatrchyan:2014tja}.
However, if we study the predictions for the invisible Higgs decay in a particular model, the situation can be rather different. 
In the scalar singlet dark matter model, there is no modification to Higgs physics at the LHC apart from a possibly large invisible decay to dark matter if allowed kinematically. 
Since also the Higgs production cross section is unaffected in this model, the invisible width modifies the signal strength of the Higgs decay to a $P_1 P_2$ 
final state in the following way:
\begin{align}
R_{P_1 P_2} = \frac{\sigma \times \text{BR}(h \rightarrow P_1 P_2)}{\sigma^\text{SM} \times \text{BR}(h \rightarrow P_1 P_2)^\text{SM}}=\frac{ \text{BR}(h \rightarrow P_1 P_2)}{\text{BR}(h \rightarrow P_1 P_2)^\text{SM}}
= \frac{\Gamma_h^\text{SM}}{\Gamma_h^\text{SM}+\Gamma_h^\text{inv}} = 1 -\text{BR}(h \rightarrow \text{inv}) \,.
\end{align}
The combined limit from the final states $W W^*,\,Z Z^*,\, \gamma\gamma,\, \bar{b}b,\, \tau^+ \tau^- $ is given by $R_\text{total}=1.17\pm 0.17$~\cite{Agashe:2014kda}. This leads to a $95\%$ confidence upper bound on the invisible Higgs branching ratio of
\begin{equation}
\label{eq:BRHiggsinvisible}
\text{BR}(h \rightarrow\text{inv})<0.16.
\end{equation}
This bound is valid for any model which only modifies the Higgs invisible branching ratio. Note that a statistically significant deviation of the combined signal strength above one, $R_\text{total}>1$, would rule out this simple dark matter model up to $M_S = M_h/2$. The bound obtained here is used when we later show the allowed parameter space in the low-mass region, 
see Figs.~\ref{fig:parameterspacefull} and \ref{fig:parameterspacelow} for details.

\subsubsection{Relic Density}
\label{sec:RelicDensity}

\begin{figure}[t]
 \includegraphics[width=0.6\linewidth]{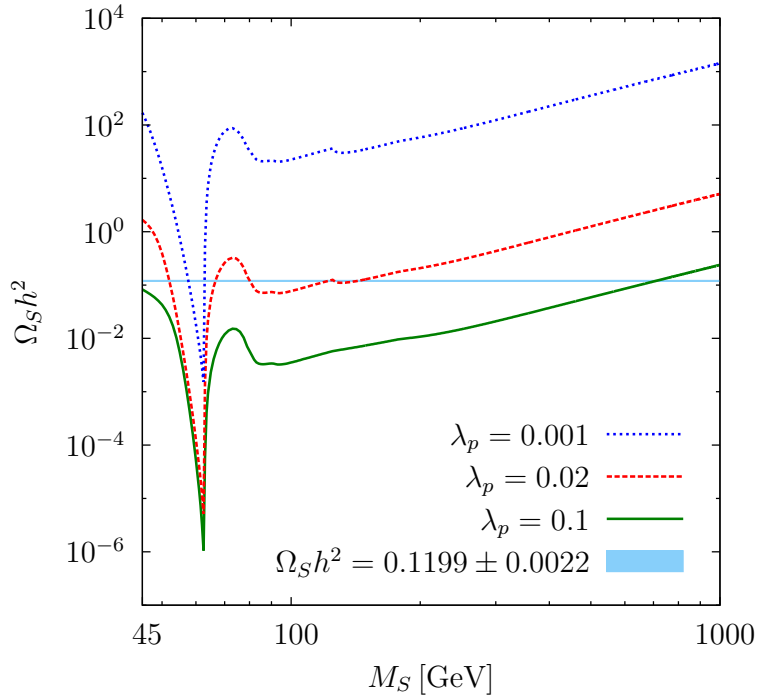}
 \caption{Dark matter relic density $\Omega_S h^2$ as a function of the dark matter mass $M_S$ for different values of the portal coupling: $\lambda_p = 0.1$~(solid green), $\lambda_p = 0.02$~(dashed red), and $\lambda_p = 0.001$~(dotted blue). The thin band where $S$ makes up the full DM relic density today, $\Omega_\text{DM} h^2 = 0.1199 \pm 0.0022$~\cite{Ade:2015xua}, is marked in light blue.}
 \label{fig:relicdensity}
\end{figure}

In order to compute the relic density of our dark matter candidate $S$, we use the analytic approximation~\cite{Gondolo:1990dk}
\begin{equation}
\label{eq:relicdensity}
\Omega_S h^2 = \frac{\unit[1.07 \times 10^{9}]{GeV}^{-1}}{J(x_f) \sqrt{g_\ast} \ M_\text{Pl}},
\end{equation}
where $M_\text{Pl}=\unit[1.22 \times 10^{19}]{GeV}$ is the Planck scale, $g_\ast$ is the total number of effective relativistic degrees of freedom at the time of freeze out, and the function $J(x_f)$ reads as
\begin{equation}
J(x_f)=\int_{x_f}^{\infty} \frac{ \langle \sigma v_\text{rel} \rangle (x)}{x^2} dx.
\end{equation}
The freeze-out parameter $x_f=M_S/T_f$ can be computed by solving
\begin{equation}
x_f= \ln \left( \frac{0.038 \ g \ M_\text{Pl} \ M_S \ \langle\sigma v_\text{rel} \rangle (x_f) }{\sqrt{g_\ast x_f}} \right),
\end{equation}
where $g$ is the number of degrees of freedom of the dark matter particle. 
Details on the calculation of the cross sections for the different DM annihilation channels and the corresponding analytic formulas, including the expressions to perform the thermal average of the cross section times relative velocity $\langle \sigma v_\text{rel}\rangle$, can be found in Appendix~\ref{app:DM}.

In Fig.~\ref{fig:relicdensity} we show the dark matter relic density as a function of the mass $M_S$, for different values of the portal coupling $\lambda_p$. 
Depending on the value of the coupling, the correct relic density can be achieved in the low-mass regime around the resonance at half of the SM Higgs mass 
$M_h = \unit[125.7]{GeV}$~\cite{Agashe:2014kda}, or off the resonance in the high-mass regime. For some values of $\lambda_p$ there is a solution in both regimes. 

\begin{figure}[t]
 \includegraphics[width=0.6\linewidth]{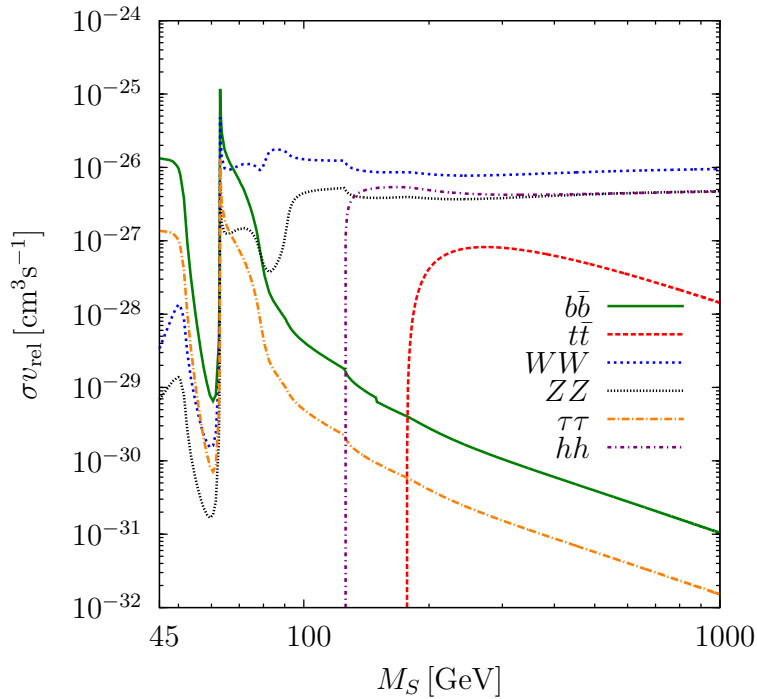}
 \caption{Cross sections times velocity $\sigma v_\text{rel}$ for the relevant dark matter annihilation channels as a function of the dark matter mass $M_S$, setting the coupling $\lambda_p$ such that we have the correct relic density for every value of the dark matter mass.} 
 \label{fig:crosssections}
\end{figure}

In order to understand the different annihilation channels relevant for our study we show in Fig.~\ref{fig:crosssections} the cross sections times velocity $\sigma v_\text{rel}$ for the different DM annihilation channels in agreement 
with the DM relic density constraints. For every value of the dark matter mass $M_S$, we use the corresponding coupling $\lambda_p$ that results in today's full DM relic density, $\Omega_\text{DM} h^2 = 0.1199 \pm 0.0022$~\cite{Ade:2015xua}. As expected, in the low mass regime the dominant channels are $b\bar{b}$ and $\tau^+ \tau^-$, and after threshold the annihilation into $W^+ W^-$ and $ZZ$ become dominant. Below $M_S = \unit[150]{GeV}$, we calculated the cross sections from the tabulated partial Higgs widths~\cite{Heinemeyer:2013tqa}, such that three- and four-body decays of the gauge bosons below threshold as well as QCD corrections are included; see Appendix~\ref{app:DM} for details. In the high-mass regime, the contributions from the annihilation into the SM Higgs $h$ and top quark pairs are significant. 

\subsubsection{Direct Detection}
To discuss the possible constraints from dark matter direct detection experiments we need to know the elastic nucleon--DM cross section. In the scalar singlet DM model, the spin-independent nucleon--DM cross section is given by
\begin{equation}
\sigma_\text{SI}=\frac{\lambda_p^2 f_N^2 \mu^2 m_N^2}{\pi M_h^4 M_S^2},
\end{equation}
where $m_N = (m_p + m_n)/2 = \unit[938.95]{MeV}$ is the 
nucleon mass for direct detection, $f_N=0.30\pm 0.03$ is the matrix element~\cite{Cline:2013gha}, and $\mu= m_N M_S/ (m_N + M_S)$ is the reduced nucleon mass. 

\begin{figure}[t]
 \includegraphics[width=0.6\linewidth]{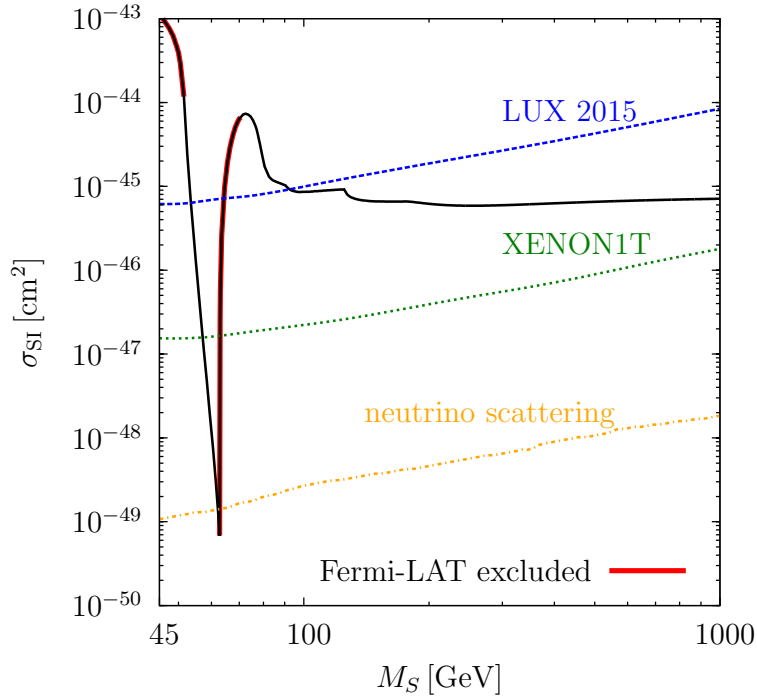}
 \caption{Spin-independent nucleon--DM cross section $\sigma_\text{SI}$. The prediction for the cross section shown here (black solid curve) is in agreement with the relic density constraints. In blue (dashed) we show the LUX bounds~\cite{Akerib:2015rjg} and in green (dotted) we show the future reach of XENON1T~\cite{Aprile:2015uzo}. In orange (dash-dotted) we show the coherent neutrino scattering background~\cite{Billard:2013qya}. The red part of the curve is excluded by the $b \bar{b}$ limits from Fermi-LAT~\cite{Ackermann:2015zua}, see Fig.~\ref{fig:indirectdetectionfull} for more details.} 
 \label{fig:directdetectionfull}
\end{figure}

In Fig.~\ref{fig:directdetectionfull} we show the predictions for the spin-independent nucleon--DM cross section $\sigma_\text{SI}$ for the typical choice $f_N = 0.30$ and the corresponding experimental bounds. 
This model is very simple and one can predict clearly the values for the elastic cross section once the relic density constraints are imposed.
As it is well known, the experimental bounds assume that the dark matter particle under study makes up 100$\%$ of the DM of the Universe. An important observation is that around the Higgs mass resonance direct detection experiments are not able to probe the parameter space in the near future as apparent in Fig.~\ref{fig:directdetectionfull}. However, as will be shown later, indirect searches are particularly sensitive to the resonant region and thus highly complementary to direct detection experiments. The projected limits by XENON1T~\cite{Aprile:2015uzo} tell us that one can test this model for a dark matter mass up to a few TeV. 

\begin{figure}[t]
 \includegraphics[width=0.6\linewidth]{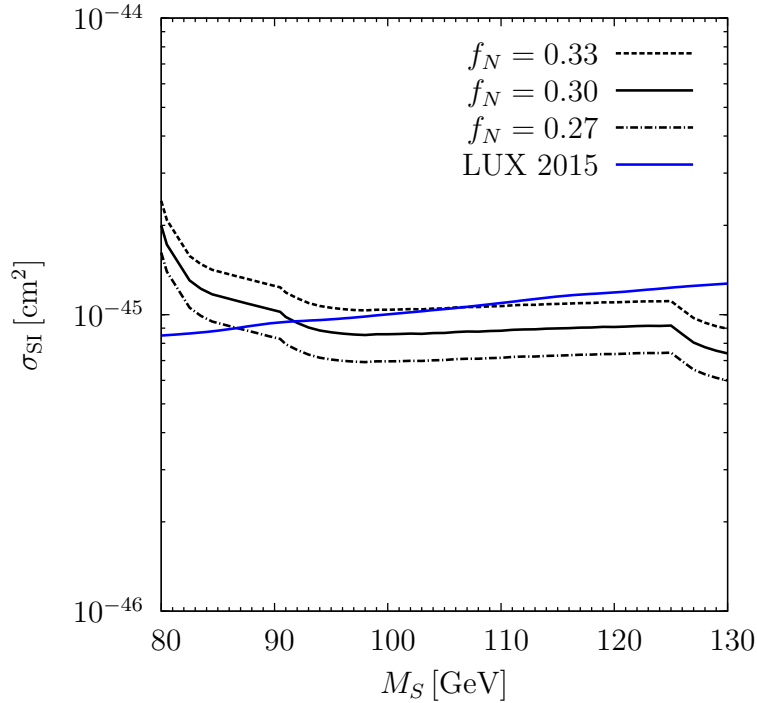}
 \caption{Spin-independent nucleon--DM cross section $\sigma_\text{SI}$ for different values of $f_N$. In blue (solid) we show the current LUX bound~\cite{Akerib:2015rjg}.} 
 \label{fig:directdetectionfN}
 \end{figure}
 
Note that the current limit from LUX~\cite{Akerib:2015rjg} on the scalar singlet DM model strongly depends on the particular value that is chosen for $f_N$. In Fig.~\ref{fig:directdetectionfN}, we show the predictions for $f_N = 0.27, \ 0.30, \ 0.33$. Depending on $f_N$, the limit on the dark matter mass $M_S$ varies between $\unit[86]{GeV}$ and $\unit[106]{GeV}$.

\subsubsection{Missing Energy Searches}
As it is well known, one can hope to observe missing energy signatures at colliders from the presence of a dark matter candidate.
We have discussed the low mass region where using the invisible decay of the Standard Model Higgs one can constrain a small part 
of the allowed parameter space in this model. Unfortunately, in the resonance region the invisible branching ratio 
of the Higgs can be very small and one cannot test this model in the near future. In the heavy mass region one can use mono-jets 
and missing energy searches where one produces the scalar singlet through the Standard Model Higgs. Unfortunately, in this case 
the production cross sections are small and this analysis is very challenging. See Ref.~\cite{Craig:2014lda} for a recent discussion. 
\subsubsection{Indirect Detection}

\begin{figure}[t]
\centering
\subfloat[]{
\includegraphics[width=0.48\linewidth]{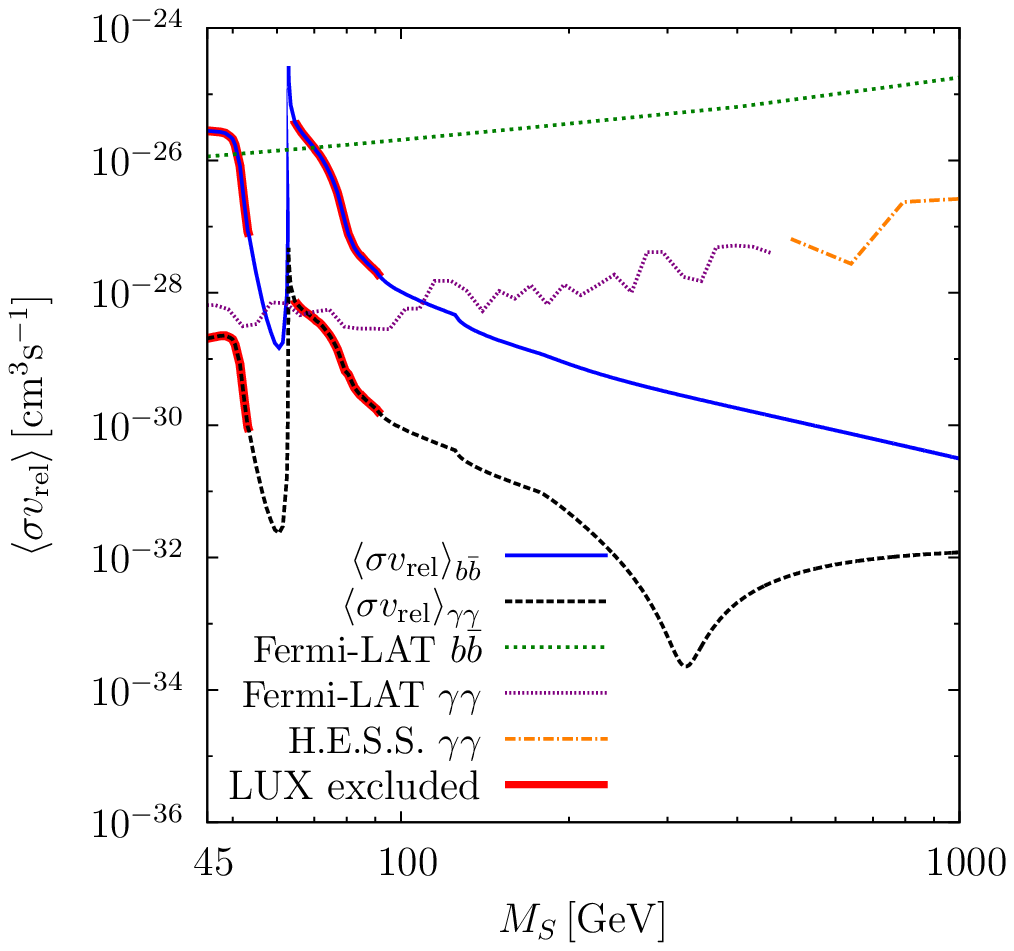}
}
\subfloat[]{
 \includegraphics[width=0.48\linewidth]{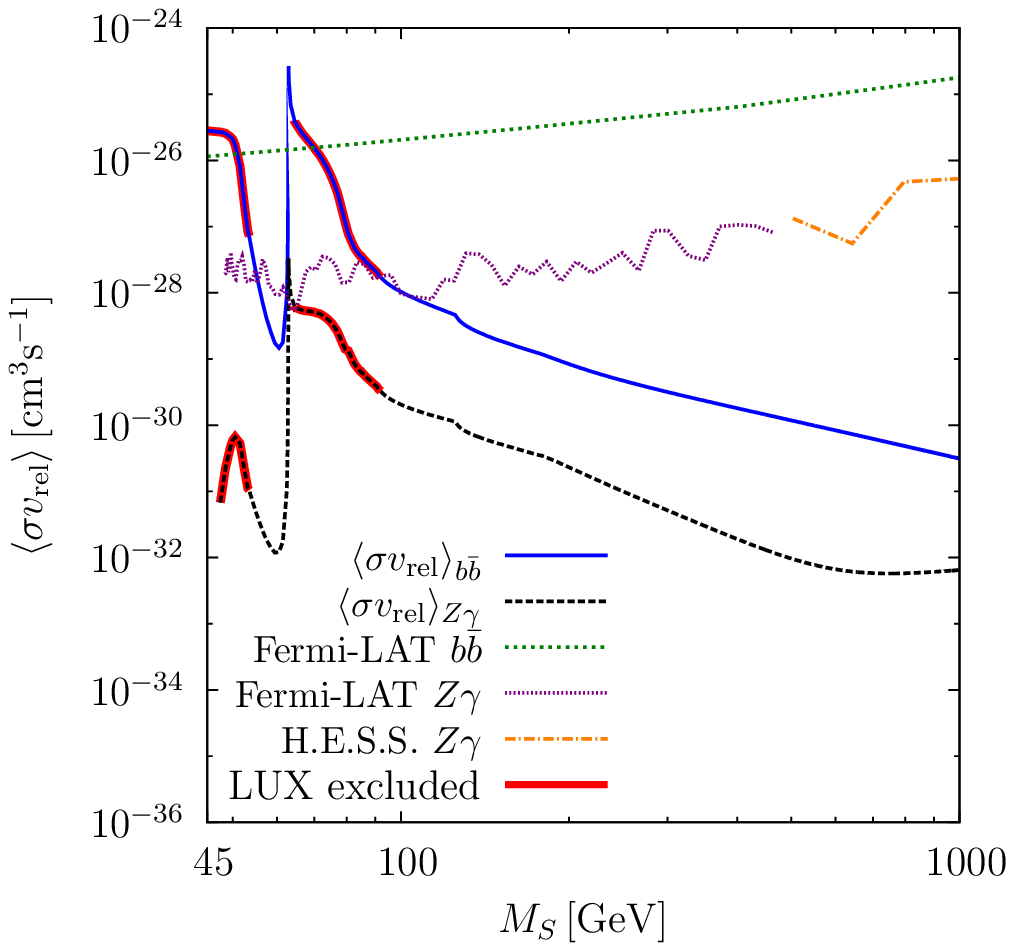}
}
\caption{Velocity-averaged cross sections times velocity for the different relevant indirect detection channels using the value of the portal coupling that gives the right relic density. We show the corresponding bounds from the Fermi-LAT~\cite{Ackermann:2015lka,Ackermann:2015zua} and H.E.S.S.\ collaborations~\cite{Abramowski:2013ax}. The red parts of the curves are excluded by the LUX direct detection limits~\cite{Akerib:2015rjg}, see Fig.~\ref{fig:directdetectionfull} for more details. (a)~Annihilation into $b \bar{b}$ and $\gamma \gamma$: the blue (solid) line shows the prediction for the annihilation into two $b$ quarks, while the black (dashed) line shows the prediction for the annihilation into $\gamma \gamma$. The dip and subsequent increase in the DM annihilation cross section to $\gamma \gamma$ is a feature of the loop functions involved in $h^\ast \to \gamma \gamma$, see Eqs.~\eqref{eq:generalAnnihilation}--\eqref{eq:loopFunctions} in Appendix~\ref{app:DM} for more details. (b)~Annihilation into $b \bar{b}$ and $Z \gamma$: the blue (solid) line shows the prediction for the annihilation into two $b$ quarks, while the black (dashed) line shows the prediction for the annihilation into $Z \gamma$. \label{fig:indirectdetectionfull}}
\end{figure}

In Fig.~\ref{fig:indirectdetectionfull} (a) and (b) we show the predictions for the dark matter annihilation into $b\bar{b}$ as well as $\gamma \gamma$ and $Z \gamma$, respectively, together with the bounds from the Fermi-LAT~\cite{Ackermann:2015lka,Ackermann:2015zua} and H.E.S.S.~\cite{Abramowski:2013ax} experiments. Continuum searches in $\bar{b} b$ and line searches constrain the same mass region, with the continuum searches giving the slightly more restrictive bounds. These bounds are very important because one can rule out part of the parameter 
space close to the resonance region. This is the only way to exclude this region because the contribution to the invisible decay of the Higgs is very small. Unfortunately, in the heavy mass region the current experimental bounds cannot exclude any of the parameter space. However, in the near future these experiments could test this simple model if the dark matter mass is close to $\unit[100]{GeV}$.

\begin{figure}[t]
 \includegraphics[width=0.6\linewidth]{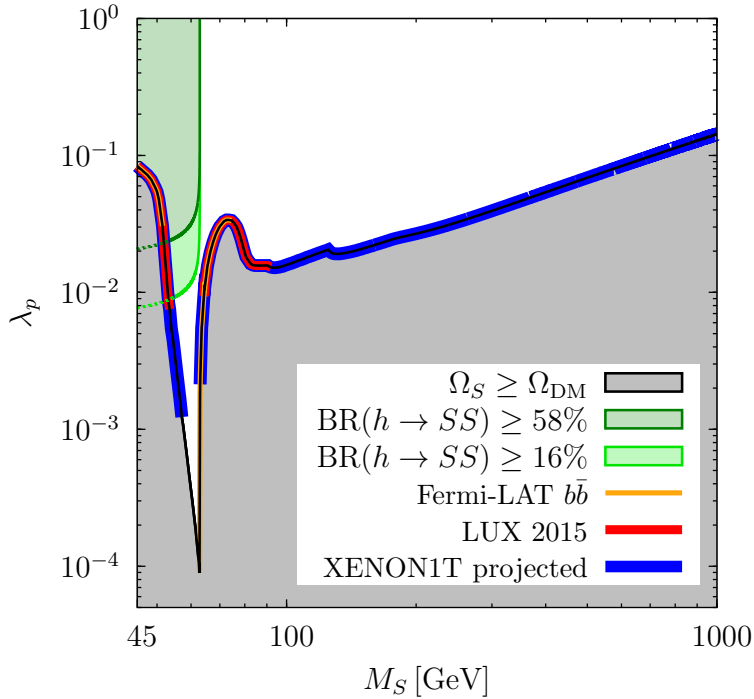}
 \caption{Allowed parameter space in the $M_S$--$\lambda_p$ plane in agreement with the relic density constraints, direct and indirect detection, and invisible Higgs decays. In gray we show the region of the parameter space where one overcloses the Universe; the black line corresponds to today's full relic density, $\Omega_\text{DM} h^2 = 0.1199\pm 0.0022$~\cite{Ade:2015xua}. In green we show the bounds from the invisible decay of the SM Higgs, using the CMS bound $\text{BR}(h \to SS) < 58\%$~\cite{Chatrchyan:2014tja}, as well as the calculated limit from Eq.~\eqref{eq:BRHiggsinvisible}. The red part of the relic density curve is excluded by the LUX direct detection experiment~\cite{Akerib:2015rjg}, while the blue part of the curve shows the projected reach of the XENON1T experiment~\cite{Aprile:2015uzo}. The orange part of the curve is excluded by the $b \bar{b}$ limits from Fermi-LAT~\cite{Ackermann:2015zua}.} 
 \label{fig:parameterspacefull}
\end{figure}

\subsubsection{Summary}

In Fig.~\ref{fig:parameterspacefull} we show the allowed parameter space in agreement with the relic density constraints, direct and indirect detection, as well as invisible Higgs decays. As one can appreciate, there are two main regions allowed by all experiments. 
In the low mass region, $\unit[53]{GeV} \leq M_S \leq \unit[62.8]{GeV}$, the dark matter annihilates through the Higgs resonance, 
while in the heavy mass region, $M_S > \unit[92]{GeV}$, all the gauge boson channels are open and dominate. In Fig.~\ref{fig:parameterspacefull} we also show the experimental bounds 
on the invisible decay of the SM Higgs~\cite{Chatrchyan:2014tja} and the projected direct detection bounds from the XENON1T experiment~\cite{Aprile:2015uzo}. The gray region is ruled out by the relic density constraints because in this region one overcloses the Universe having too much dark matter relic density. Notice that even this simple model for dark matter is not very constrained by the experiments.

\noindent \textbf{Low Mass Regime}

\begin{figure}[t]
 \includegraphics[width=0.6\linewidth]{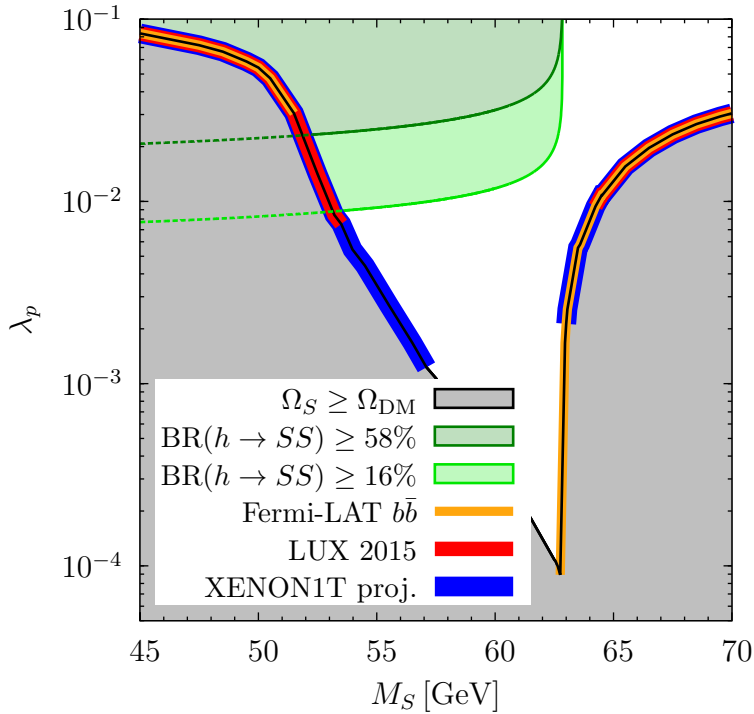}
 \caption{Allowed parameter space in the $M_S$--$\lambda_p$ plane in agreement with the relic density constraints, direct and indirect detection, and invisible Higgs decays in the low mass regime.  In this region the constraints from the invisible decay of the Higgs shown in green are very important. Color coding is the same as in Fig.~\ref{fig:parameterspacefull}.} 
 \label{fig:parameterspacelow}
\end{figure}

In the low mass region the allowed dark matter mass is $\unit[53]{GeV} \leq M_S \leq \unit[62.8]{GeV}$. In this region close to the Higgs resonance the dark matter can annihilate into Standard Model fermions 
or into two fermions and a gauge boson. 
In Fig.~\ref{fig:parameterspacelow} we show a detailed analysis 
of this region to understand which part of the parameter space is ruled out by experiments. Notice that the main annihilation 
channel is $SS \to \bar{b} b$. In this model one can set 
bounds only using the constraints on the nucleon--DM cross section. The scattering 
between electrons and DM is highly suppressed by the small Yukawa coupling.

From the results presented in Fig.~\ref{fig:parameterspacelow} one can see that the resonance region cannot be excluded 
or tested in the near future by direct detection experiments. This is a pessimistic result, but fortunately this region can be tested 
at gamma-ray telescopes as we will discuss in the next section. 

\noindent \textbf{Heavy Mass Regime}

When the dark matter is heavy it can annihilate into all Standard Model particles. 
For $M_S \geq \unit[70]{GeV}$, where the dominant annihilation channels are into gauge bosons, the current direct detection bound from LUX~\cite{Akerib:2015rjg} rules out only a very small range in the parameter space, 
$M_S \lesssim \unit[92]{GeV}$, see Fig.~\ref{fig:parameterspacefull}. It means that this model is not very constrained by direct detection in the heavy region. These results are crucial to understand 
the testability of this simple model at gamma-ray telescopes.

\section{Dark Matter Annihilation into Gamma-Ray Lines}
\label{sec:DMGammaLines}
The detection of a monochromatic gamma line coming from dark matter annihilation would be a very strong hint towards the particle interpretation of the dark matter in the Universe. 
It is highly unlikely that an astrophysical compact source would generate very energetic monochromatic photons in different regions of the galaxy. 
The question whether a gamma line signal is generic and visible in a dark matter model is very subtle. Whenever tree-level annihilations into SM particles are present, 
one expects that the final state radiation off the charged SM particles can make the gamma line undetectable experimentally. 
To investigate the line visibility in the scalar singlet dark matter model we compute the cross sections for the final state radiation processes and investigate 
the gamma flux spectra in the relevant parts of the parameter space. 

\subsection{Final State Radiation}
There are three relevant regions which define the properties of the gamma spectrum coming from dark matter annihilation. 
Let us define $x_\gamma = E_\gamma/M_S$, where $E_\gamma$ is the energy of the photon and $M_S$ is the dark matter mass. 
When $x_\gamma$ is very small one has the photons coming mainly from hadronization, i.e., the dark matter annihilates into quarks 
and from the cascade one has the photons with a continuous spectrum. When $x_\gamma$ is close to one, one finds that the final 
state radiation processes contribute more because they can provide hard photons. Finally, when $x_\gamma = 1$ one has
the gamma line with energy equal to the DM mass. Therefore, one must understand the final state radiation processes to investigate the visibility of the gamma lines.

The relevant final state radiation process for our study is $ S S \to \bar{X}X \gamma $, with the kinematic endpoint of the continuous $\gamma$ spectrum at
\begin{equation}
 E_\gamma^\text{max} = M_S \left( 1 - \frac{M_X^2}{M_S^2} \right),
\end{equation}
in the non-relativistic limit $s= 4 M_S^2$. In the low-mass regime the dominant process is  $S S \to\bar{f}f \gamma $ with the strongest contribution from the bottom quark, while in the high-mass regime  $ S S \to W^+W^-\gamma$ becomes dominant. The differential cross section times velocity of those processes is given by
\begin{equation}\label{eq:FinalStateRadiation}
 \frac{d \sigma v_\text{rel}}{dE_\gamma dE_1} = \frac{1}{32 \pi^3 s} \left| \mathcal{M}_\text{FSR} \right|^2,
\end{equation}
where the integration limits for the integration over $E_1$ for a fixed $E_\gamma$ are given by
\begin{align}
 E_1^\text{min} &= M_S - \frac{E_\gamma}{2} - \frac{\sqrt{ E_\gamma^2 (E_\gamma - M_S) M_S \left[M_X^2 + (E_\gamma - M_S) M_S \right]}}{2 M_S (M_S - E_\gamma)}, \\
 E_1^\text{max} &=M_S - \frac{E_\gamma}{2} + \frac{\sqrt{ E_\gamma^2 (E_\gamma - M_S) M_S \left[M_X^2 + (E_\gamma - M_S) M_S \right]}}{2 M_S (M_S - E_\gamma)}, 
\end{align}
in the limit  $s= 4 M_S^2$. See Appendix~\ref{app:FSR} for the amplitudes of the two relevant processes for final state radiation, $SS \to \bar{f} f \gamma$ and $SS \to W^+W^-\gamma$. Notice that in the low-mass region the FSR is suppressed by small Yukawa couplings. Therefore, this is the region where generically one can have a visible gamma line. 

\subsection{Gamma Flux}

The differential photon flux is given by 
\begin{equation}
\frac{d \Phi_\gamma}{dE_\gamma} = \frac{n_\gamma}{8 \pi \, M_S^2} J_\text{ann} \frac{ d \langle \sigma v_\text{rel} \rangle}{dE_\gamma} = \frac{n_\gamma}{8 \pi \, M_S^2} J_\text{ann} \langle \sigma v_\text{rel} \rangle \frac{ dN_\gamma}{dE_\gamma}\, ,
\end{equation}
where the factor $J_\text{ann}$ contains the astrophysical assumptions about the DM distribution in the galaxy and thus all the astrophysical uncertainties. 
Here $n_\gamma$ is the number of photons per annihilation, and $d N_\gamma/d E_\gamma$ is the differential energy spectrum of the photons coming from dark matter annihilation.
In all numerical calculations, we will use the $J$-factor from the R3 region-of-interest, given by the Fermi-LAT collaboration to be $J_\text{ann} = \unit[13.9\times 10^{22}]{GeV^2 cm^{-5}}$~\cite{Ackermann:2013uma}. 
The R3 region is a circular region of radius $3^\circ$ centered on the galactic center~\cite{Ackermann:2013uma}.
The differential flux of the line is extremely narrow, however to make connection with the experiment it will be folded with a Gaussian function modeling the detector resolution.
\begin{itemize}
\item $ S  S \rightarrow  \gamma \gamma$: for the annihilation into two photons the flux is given by
\begin{equation}
\frac{d \Phi_\gamma}{d E_\gamma} =  \frac{  J_\text{ann} \langle \sigma v_\text{rel} \rangle_{\gamma \gamma}}{4 \pi M_S^2 } \int_{0}^\infty dE_0 \ \delta (E_0 - M_S) \ G(E_\gamma, \xi/w, E_0) \,,
\end{equation}
with 
\begin{equation}
 G(E_\gamma, \xi/w, E_0) = \frac{ \exp^{-\frac{ (E_\gamma -E_0)^2}{2 E_0^2 (\xi/w)^2}}}{\sqrt{2 \pi }
   E_0  (\xi/w) }\,.
\end{equation}
The parameter $\xi$ is a measure of the detector energy resolution which varies between $0.01$ and $0.1$ in the relevant energy range. The factor $w = 2 \sqrt{2 \log{2}} \approx 2.35$ determines the full width at half maximum as $ \sigma_0 w = \xi E_0$, therefore we have $\sigma_0 =  E_0\xi/w$ in the usual Gaussian function. For the annihilation to $\gamma\gamma$, the energy of the gamma line is at the dark matter mass,
\begin{equation}
 E_\gamma = M_S.
\end{equation}

\item $S S  \rightarrow  X \gamma$: for the annihilation into an unstable final state particle along with a photon, the flux is given by 
\begin{equation}
\label{eq:annihilationXgamma}
\frac{d \Phi_\gamma}{d E_\gamma} =  \frac{ J_\text{ann} \langle \sigma v_\text{rel} \rangle_{X \gamma} }{8 \pi M_S^2 } \int_{0}^\infty dE_0 \frac{1}{\pi}\frac{4  M_S M_X \Gamma_X}{(4M_S^2 - 4 M_S E_0 - M_X^2)^2 + \Gamma_X^2 M_X^2}  G(E_\gamma, \xi/w, E_0).
\end{equation}
Here $\Gamma_X$ is the decay width of the unstable particle in the final state and $M_X$ is its mass. See Appendix~\ref{app:gammaspectrum} for a derivation of the differential energy spectrum used in Eq.~\eqref{eq:annihilationXgamma}. The gamma line energy is given by 
\begin{equation}
 E_\gamma = M_S \left(1 - \frac{M_X^2}{4 M_S^2}\right).
\end{equation}
\end{itemize}
Using these expressions for the differential flux we now study the predictions for the gamma lines in this model in the benchmark scenarios defined in Table~\ref{tab:benchmarkscenarios}.

\begin{table}[t]
\centering
\caption{Benchmark scenarios for the study of the $\gamma$ spectrum. \label{tab:benchmarkscenarios}}
\begin{tabular}{cccc}
\hline \hline
 Scenario & $M_S$~[GeV] & $\lambda_p$ & Energy of the $Z\gamma$ line~[GeV] \\ 
\hline 
 1 & 62.5 & $9.06\times 10^{-5}$ & 29.2 \\ 
 2 & 150  & $2.08\times 10^{-2}$ & 136 \\ 
 3 & 316  & $4.17\times 10^{-2}$ & 309 \\ 
 4 & 500  & $6.87\times 10^{-2}$ & 496 \\ 
 \hline \hline
\end{tabular}
\end{table}

\begin{figure}
\centering
\subfloat[]{
 \includegraphics[width=0.48\linewidth]{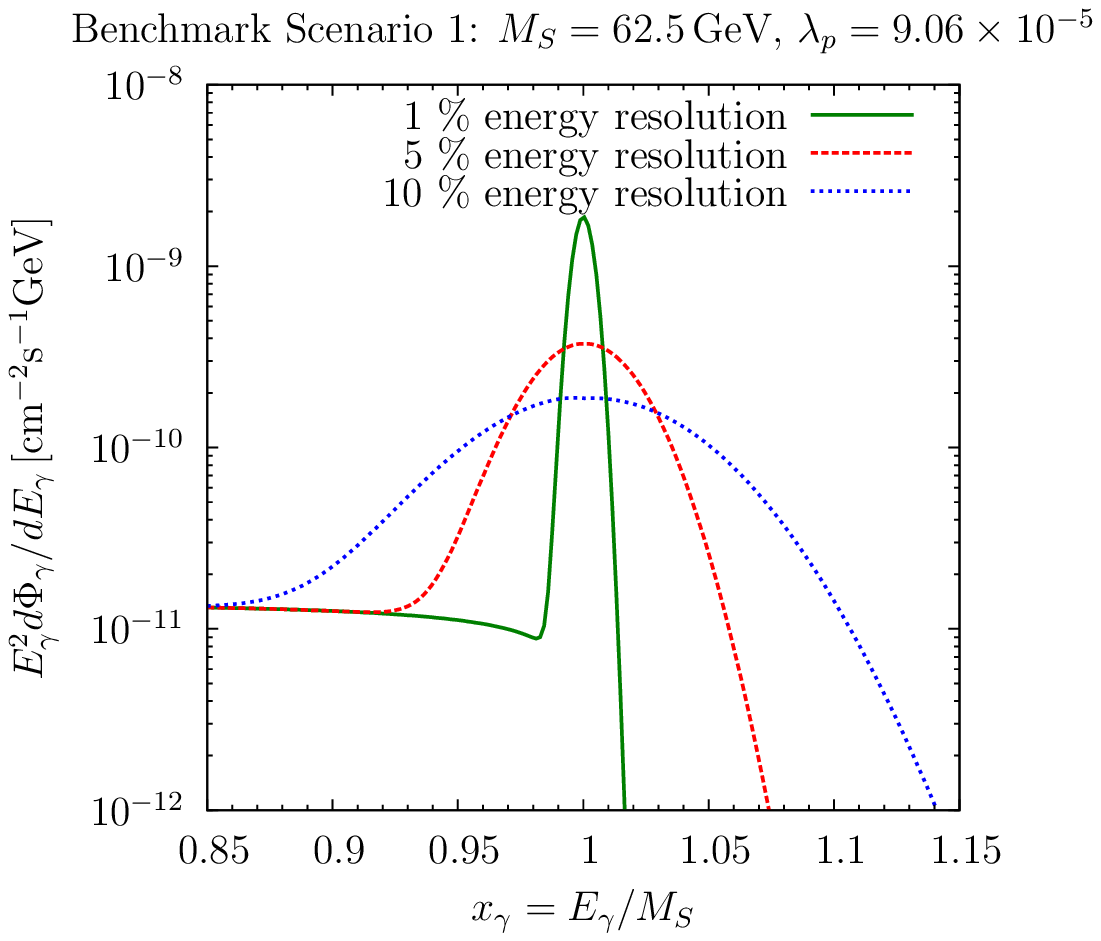}
}
\subfloat[]{
 \includegraphics[width=0.48\linewidth]{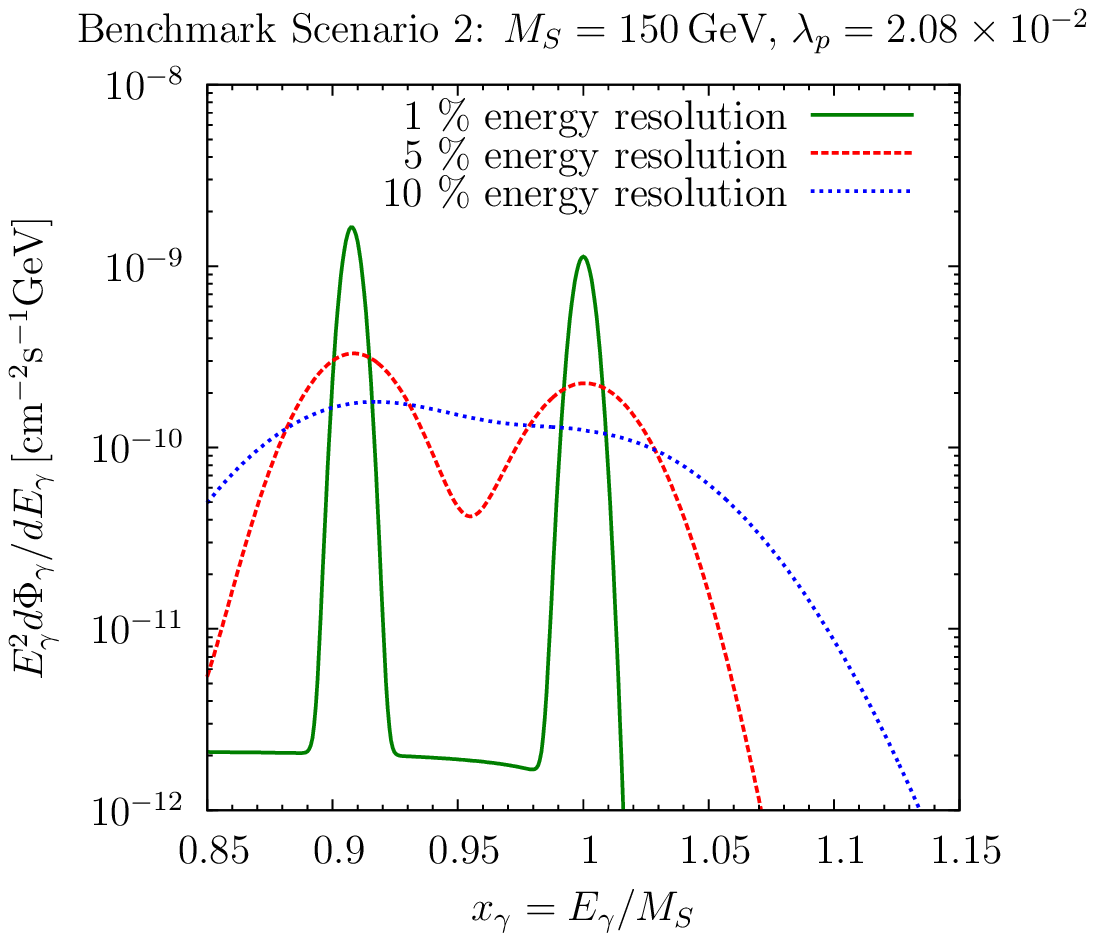}
}
\vspace{2mm}
\subfloat[]{
 \includegraphics[width=0.48\linewidth]{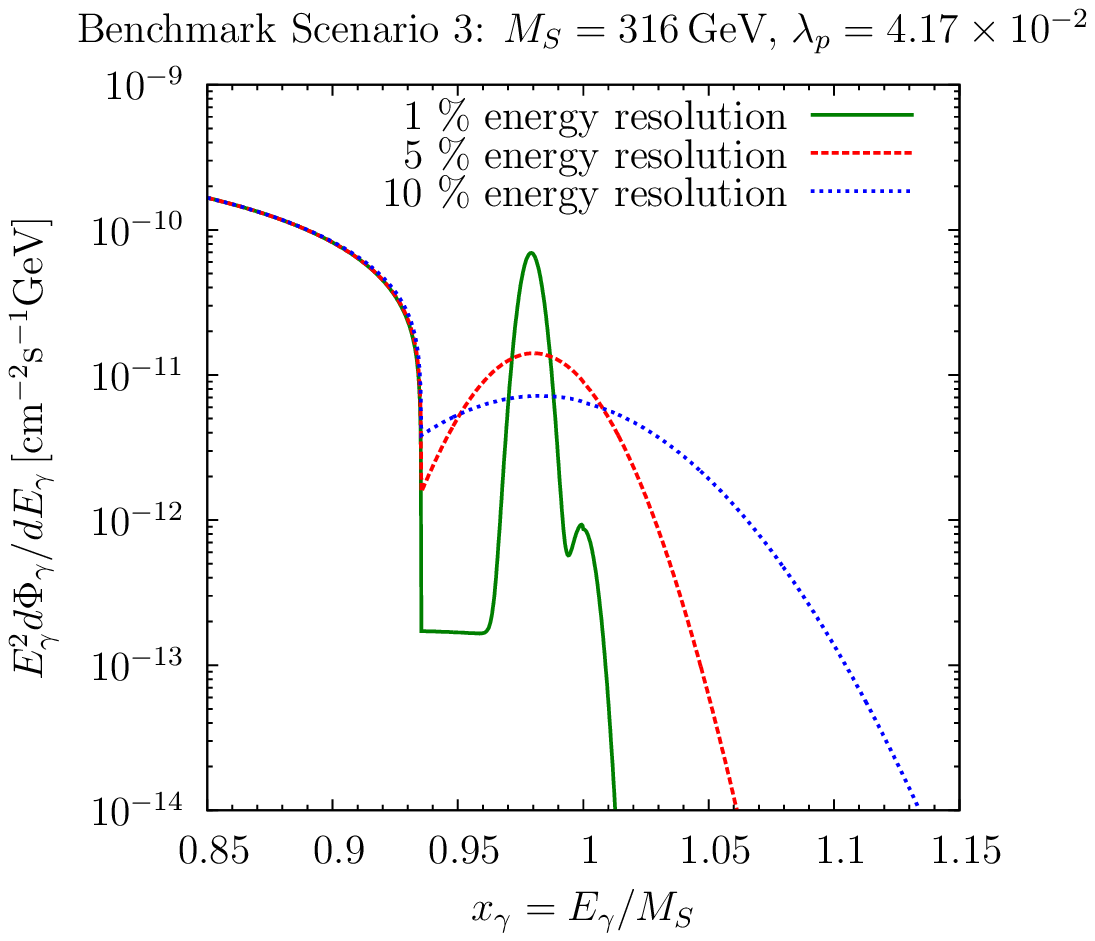}
}
\subfloat[]{
 \includegraphics[width=0.48\linewidth]{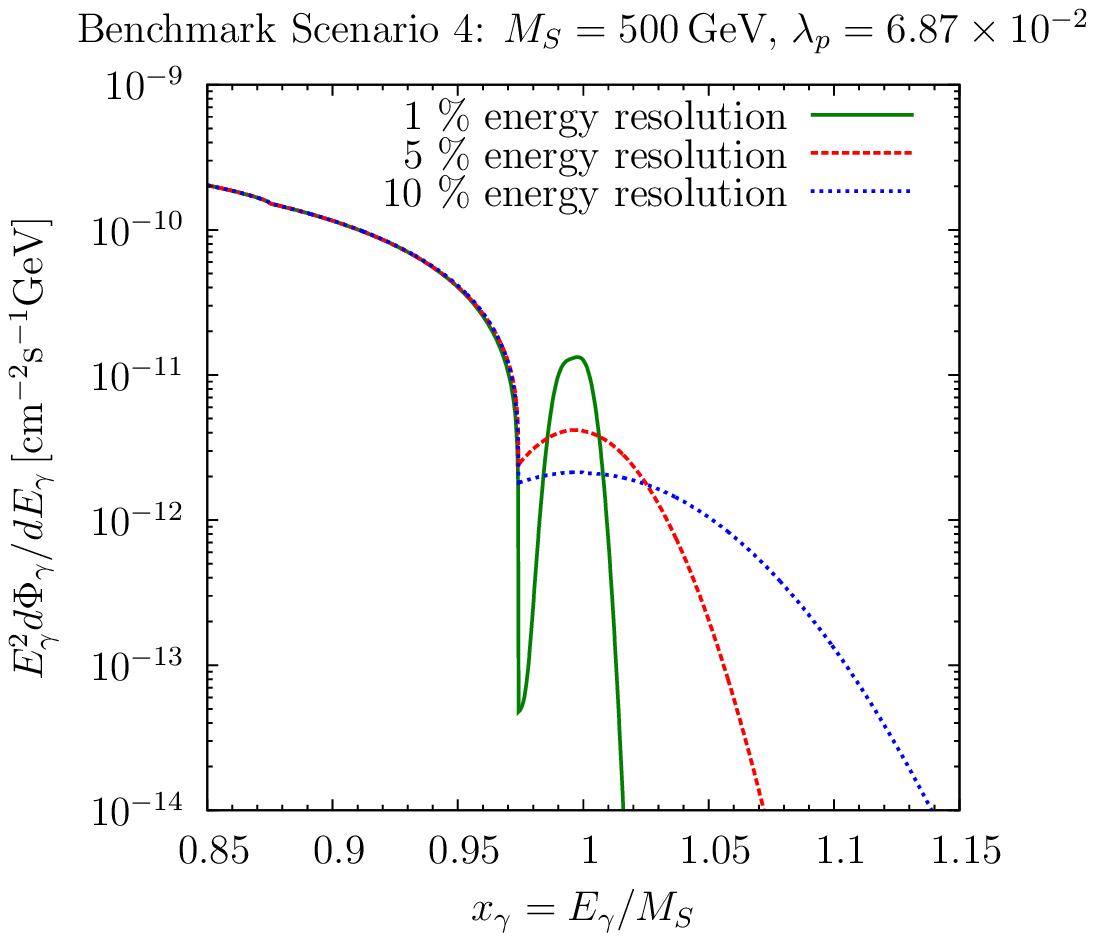}
}
\caption{Spectra for the benchmark scenarios from Tab.~\ref{tab:benchmarkscenarios}. The different curves correspond to an energy resolution of $1\%$ (solid green), $5\%$ (dashed red) and $10\%$ (dotted blue), respectively. (a)~Benchmark scenario 1: the main contribution to final state radiation is the annihilation to $b\bar{b}\gamma$. (b)~Benchmark scenario 2: the main contribution to final state radiation is the annihilation to $WW\gamma$. (c)~Benchmark scenario 3: the main contribution to final state radiation is the annihilation to $WW\gamma$. (d)~Benchmark scenario 4: the main contribution to final state radiation is the annihilation to $WW\gamma$. \label{fig:spectrum}}
\end{figure}

In Fig.~\ref{fig:spectrum}~(a) we show the gamma spectrum for the scenario 1 with $M_S=\unit[62.5]{GeV}$. In this case one has the 
resonant dark matter annihilation through the SM Higgs. As one can see in this scenario it is possible to identify 
the gamma line from DM annihilation into $\gamma \gamma$, while the line from $Z \gamma$ is not visible in the plot since it is at $x_\gamma = 0.47$ and will be swamped in the FSR background. 
The main contribution to final state radiation in this case is coming from the annihilation into $\bar{b} b \gamma$ 
but it is suppressed by the small bottom Yukawa coupling. Therefore, in this case one has a large difference 
between the final state radiation and the gamma line.

In Fig.~\ref{fig:spectrum}~(b) the predictions for the gamma spectrum is shown for the second scenario where $M_S=\unit[150]{GeV}$.
This scenario is ideal because one can see the two possible lines in this model, the $\gamma \gamma$ and $Z \gamma$ 
lines, if one has a good energy resolution. In this case the main contribution to final state radiation comes from the DM 
annihilation to $WW\gamma$. However, as one can appreciate, there is a large difference between FSR and the gamma lines because the endpoint of the FSR is far from the DM mass.

The case when the DM mass is $\unit[316]{GeV}$ is shown in Fig.~\ref{fig:spectrum}~(c). There is a large difference between the rate for the $Z\gamma$ 
and $\gamma \gamma$ lines. Unfortunately, in this case one could see the lines only with a perfect energy resolution. 
The cross section for the final state radiation processes is large in this case making the observation of gamma lines very challenging.
Finally, we present in Fig.~\ref{fig:spectrum}~(d) the energy spectrum for the case when $M_S=\unit[500]{GeV}$. In this case one cannot distinguish the gamma lines since the difference between the final state radiation and the gamma line is very small.  

\begin{figure}[t]
\centering
\subfloat[]{
\includegraphics[width=0.48\linewidth]{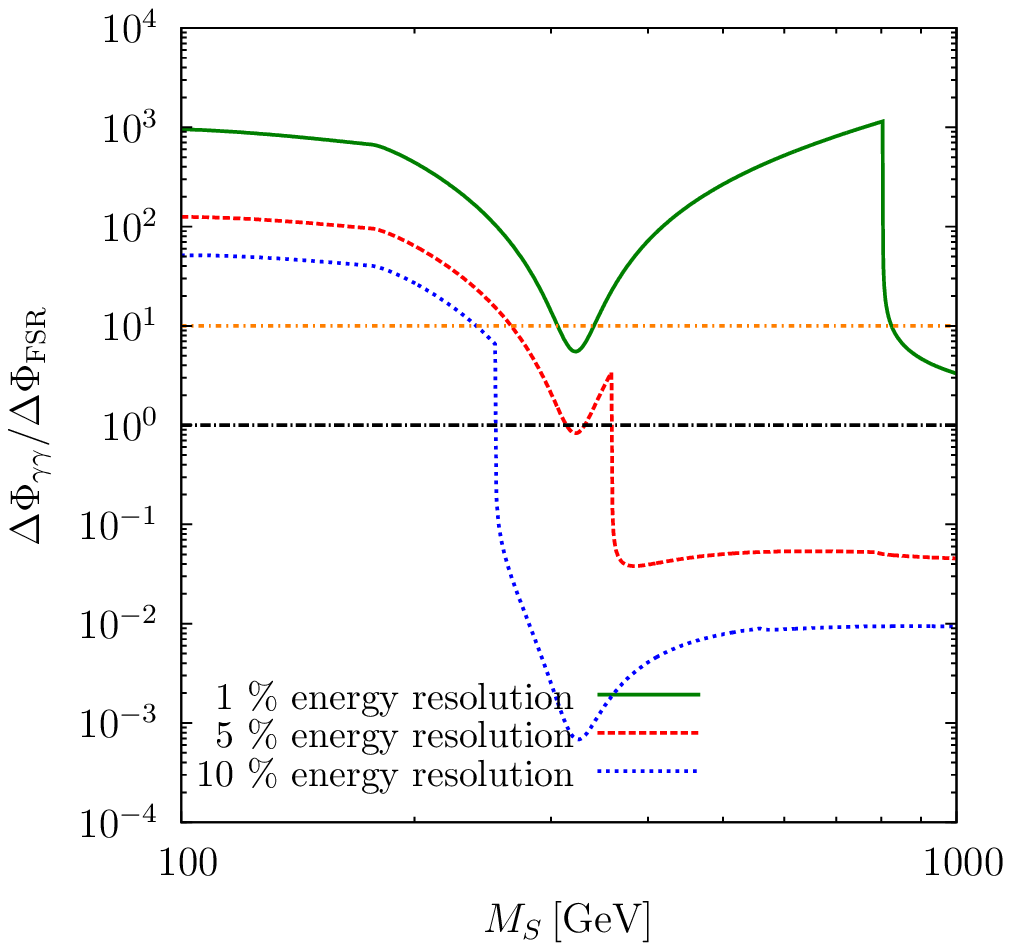}
}
\subfloat[]{
 \includegraphics[width=0.48\linewidth]{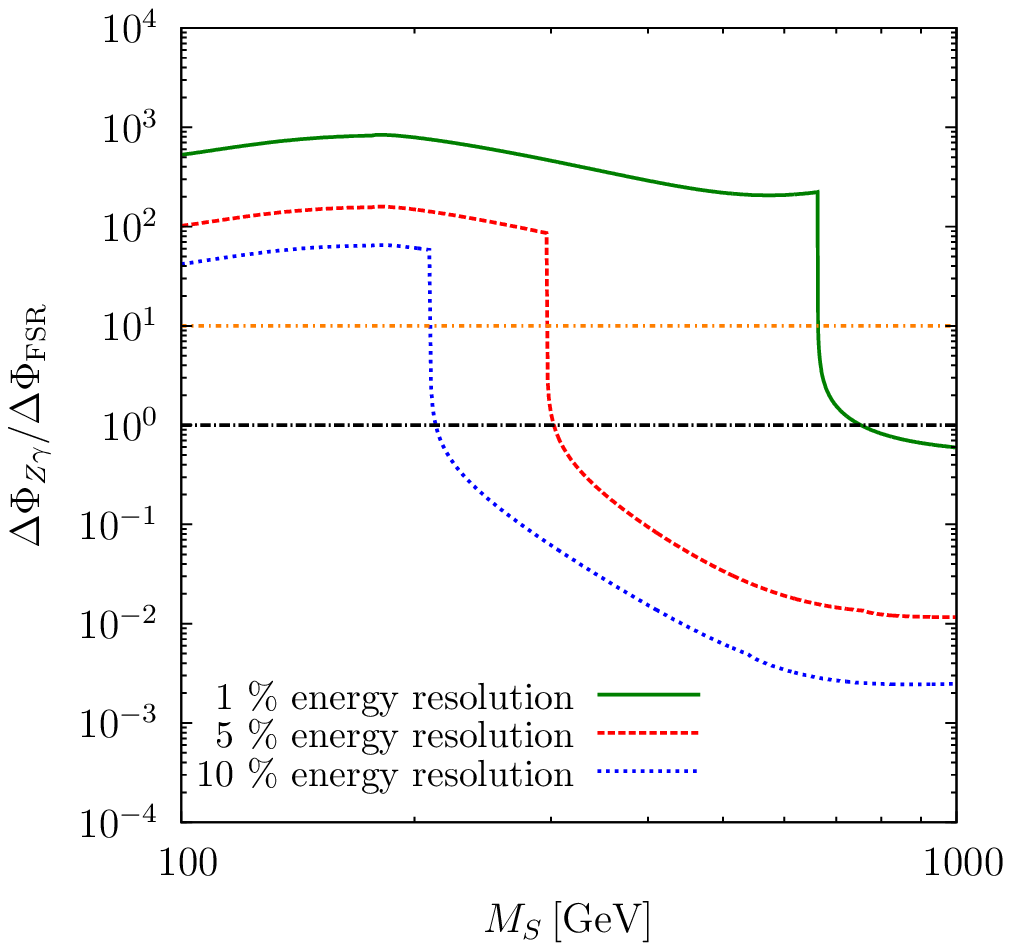}
}
\caption{Line visibility for the three different experimental energy resolutions used before: $1\%$ (solid green), $5\%$ (dashed red) and $10 \%$ (dotted blue). To estimate the visibility of the line, we mark the ratio 1 in black (dash-dotted) and the ratio 10 in orange (dash-dotted). (a)~Visibility of the $\gamma \gamma$ line: we show the ratio of the differential gamma flux at the line energy to the FSR flux at 90 \% of the dark matter mass for 10 \% energy resolution, at 95 \% for 5 \% energy resolution, and at 99 \% for 1 \% energy resolution. (b)~Visibility of the $Z \gamma$ line: we show the ratio of the differential gamma flux at the line energy to the FSR flux at 90 \% of the line energy for 10 \% energy resolution, at 95 \% for 5 \% energy resolution, and at 99 \% for 1 \% energy resolution. \label{fig:visibility}}
\end{figure}

In order to have a more generic discussion about the visibility of the gamma lines in Fig.~\ref{fig:visibility} 
we show the ratios between the $\gamma \gamma$ and $Z\gamma$ fluxes and the gamma flux from final state radiation. We display these ratios
for $M_S \geq \unit[100]{GeV}$. We show the curves for 1\% energy resolution (green solid), 5\% energy resolution (red dashed), and 10\% energy resolution (blue dotted). 
To be conservative we can say that the lines are visible if the ratio between the fluxes in Fig.~\ref{fig:visibility} is larger than a factor 10. 
This means that for realistic experiments with an energy resolution of 5 \%, the gamma lines can be visible when the dark matter mass is smaller than $\unit[300]{GeV}$. 

As one can appreciate from the above discussion one could observe the gamma lines from dark matter annihilation 
in this model only when the dark matter mass is small, i.e., in the low mass region where $M_S=\unit[(53-62.8)]{GeV}$ or in the intermediate region, $M_S=\unit[(92-300)]{GeV}$, 
where the two gamma lines could be distinguished from the continuum and among each other.
These results are crucial to understand the testability of this model at gamma-ray telescopes.

In Fig.~5 one can see that the Fermi-LAT experiment could test this model in the low mass region where the gamma gamma cross sections are large. The proposed GAMMA-400 experiment~\cite{Topchiev:2015wva} will have a better energy resolution and therefore will explore the same region with a better chance to see the gamma lines. For the high-mass region, there is currently no planned experiment that will have the required sensitivity in cross section to reach the values predicted in this model. 

Now, let us make a short summary of all constraints to understand how this model can be tested in the near future. 
In Fig.~\ref{fig:directdetectionfull} we have seen that the low mass region cannot fully be tested at direct detection experiments. 
However, the region where the dark matter mass is in the range $M_S=\unit[(92-300)]{GeV}$ could be tested at the XENON1T experiment.
Notice that the spin-independent cross section for $M_S=\unit[100]{GeV}$ is $\sigma_{\rm{SI}}=\unit[8.6 \times 10^{-46}]{cm^2}$, while for 
$M_S=\unit[300]{GeV}$ one finds $\sigma_{\rm{SI}}=\unit[6 \times 10^{-46}]{cm^2}$.
At the same time in this region, one could see the gamma lines coming from dark matter annihilation as we have shown in Fig.~\ref{fig:visibility}.
The annihilation cross sections are $\langle \sigma v_\text{rel} \rangle_{\gamma \gamma}=\unit[9.1 \times 10^{-31}]{cm^3 s^{-1}}$ when $M_S=\unit[100]{GeV}$, and 
$\langle\sigma v_\text{rel} \rangle_{Z \gamma}=\unit[5.2 \times 10^{-32}]{cm^3 s^{-1}}$ for $M_S=\unit[300]{GeV}$. 
Therefore, one could say that only the intermediate region with  $M_S=\unit[(92-300)]{GeV}$ can be tested at both direct and indirect experiments. Of course, due to the smallness of the gamma line cross sections, it will be challenging to reach the required experimental sensitivity at indirect detection experiments to test this model at $M_S = \unit[300]{GeV}$.

\section{Summary}
\label{sec:Summary}
In this article we have investigated in great detail the predictions in the simplest dark matter model where the dark matter candidate is a real scalar field.
This model can be considered as a toy model for dark matter but it offers the possibility to connect the predictions for all relevant dark matter experiments.
This model has only two free parameters which are constrained by the relic density. Once one uses the relic density constraints we can make 
predictions for the elastic nucleon--DM scattering and the cross sections for DM annihilation into gamma rays. We have revisited the model and updated all constraints for the full parameter space of the model. 

We have computed for the first time all contributions to final state radiation analytically in this model. In the low mass region, $M_S=\unit[(53-62.8)]{GeV}$,  
the main contribution to final state radiation comes from the annihilation into two $b$-quarks and a photon, while in the heavy mass region 
the annihilation into two $W$ gauge bosons and a photon is the dominant contribution. These results are very important to understand the visibility of the gamma lines.

We have shown the predictions for the two possible gamma lines in this model coming from the DM annihilation into two photons and into $Z \gamma$.
In Section~\ref{sec:DMGammaLines} we have shown the numerical predictions for the gamma spectrum in different scenarios. We have shown that in the low mass region one 
could see the $\gamma \gamma$ line because the annihilation into $\bar{b}b\gamma$ is suppressed. In the intermediate region where the dark matter mass 
is  $M_S=\unit[(92-300)]{GeV}$ there is a large difference between the continuum and the gamma line and one could see the two gamma lines if the energy resolution is good. 
Unfortunately, in the heavy mass region the dark matter annihilation into $WW\gamma$ is large and it is very challenging to observe the gamma lines.
We have shown that only the region where $M_S=\unit[(92-300)]{GeV}$ can be tested at direct and indirect detection experiments. 
These results can motivate a dedicated study in this region of the parameter space if one sticks to the simplest model for dark matter.

\section*{Acknowledgments}
P.F.P.\ thanks Clifford Cheung and Mark B.\ Wise for discussions.
The work of P.F.P.\ is partially funded by the Gordon and Betty Moore Foundation through Grant 776 
to the Caltech Moore Center for Theoretical Cosmology and Physics and Walter Burke Institute for Theoretical Physics, Caltech, Pasadena CA.
P.F.P.\ thanks the theory group at Caltech for hospitality.

\appendix
\section{Invisible Higgs Decay}
In the scalar singlet DM model the Standard Model Higgs can decay into dark matter in the low mass region.
The invisible decay width of the Higgs in this case is given by
\begin{equation}
\Gamma (h \to SS) = \frac{\lambda_p^2 v_0^2}{8 \pi M_h^2} (M_h^2-4 M_S^2)^{1/2}.
\end{equation}
To calculate the invisible branching fraction
\begin{equation}
 \text{BR}(h \to SS) = \frac{\Gamma (h \to SS)}{\Gamma_h^\text{SM} +\Gamma (h \to SS) },
\end{equation}
we use the SM Higgs width $\Gamma_h^\text{SM} = \unit[4.17]{MeV}$ for $M_h = \unit[125.7]{GeV}$~\cite{Heinemeyer:2013tqa}.
\section{Dark Matter Annihilation Cross Sections}
\label{app:DM}
%
\begin{itemize}
 \item Annihilation cross section times velocity from the Higgs width:
 to obtain an accurate value for the total DM annihilation cross section to SM final states, one needs to take into account QCD corrections for quarks in the final state, as well as three- and four-body final states from virtual gauge boson decays below threshold. This is most easily done by using the tabulated Higgs width\footnote{Or the partial widths for each final state when we are interested in the individual cross sections times velocity.} as a function of the invariant mass $\Gamma_h(\sqrt{s})$~\cite{Heinemeyer:2013tqa} and rewrite the cross section times velocity as 
 \begin{equation}
 \label{eq:crossSectionWidth}
  \sigma v_\text{rel} = \frac{8 \lambda_p^2 v_0^2}{\sqrt{s}} \left| D_h(s)\right|^2 \Gamma_h(\sqrt{s}),
 \end{equation}
 with
 \begin{equation}
  \left| D_h(s)\right|^2 = \frac{1}{(s- M_h^2)^2 + M_h^2 \Gamma_h^2(M_h)}. 
 \end{equation}
 This factorization is possible for all final states except the SM Higgs, such that for $M_S > M_h$ the contribution $SS \to hh$ has to be added. For $M_S < M_h/2$, the width in $D_h(s)$ has to take into account the invisible decay $h \to SS$. The thermally averaged annihilation cross section times velocity $\langle \sigma v_\text{rel} \rangle$ is a function of $x=M_S/T$, and -- when using the expression for $\sigma v_\text{rel}$ in Eq.~\eqref{eq:crossSectionWidth} --  can be computed as
 \begin{equation}
  \langle \sigma v_\text{rel} \rangle (x) = \frac{ x }{16 M_S^5 K_2^2(x)}  \int_{4 M_S^2}^{\infty} ds  \sqrt{s-4M_S^2} \ s \ K_1\left( \frac{x \sqrt{s}}{M_S} \right) \sigma v_\text{rel},
 \end{equation}
 where $K_1$ and $K_2$ are modified Bessel functions of the second kind.

  For $M_S \geq \unit[150]{GeV}$ we use the tree-level expressions calculated below, since then loop corrections overestimate the tabulated width. See the discussion in Ref.~\cite{Cline:2013gha} for more details.
 From the cross section $\sigma$, the thermal average can be computed via
\begin{equation}
 \langle\sigma v_\text{rel} \rangle (x) = \frac{x}{8 M_S^5 K_2^2(x)} \int_{4 M_S^2}^\infty  ds  \ ( s - 4 M_S^2) \ \sqrt{s} \ K_1 \left(\frac{x \sqrt{s}}{M_S}\right) \sigma .
\end{equation}
 
 \item Annihilation into Standard Model fermions:
 \begin{equation}
\sigma ( SS \to \bar{f} f ) = \frac{\lambda_p^2 M_f^2 N_c^f (s-4 M_f^2)^{3/2}}{2 \pi s \sqrt{s-4 M_S^2} \left[ \left(s-M_h^2\right)^2 + \Gamma_h^2  M_h^2 \right] },
\end{equation}
where $N_c^f$ is the color factor of the fermion $f$.
\item Annihilation into two $W$ gauge bosons:
 \begin{equation}
  \sigma ( SS \to WW) = \frac{\lambda_p^2}{4 \pi s} \frac{\sqrt{s - 4 M_W^2}}{\sqrt{s - 4 M_S^2}} \frac{\left(s^2 - 4 M_W^2 s + 12 M_W^4\right)}{\left[ \left( s - M_h^2 \right)^2 + \Gamma_h^2 M_h^2 \right]}.
 \end{equation}
 \item Annihilation into two $Z$ gauge bosons:
  \begin{equation}
  \sigma ( SS \to ZZ) = \frac{\lambda_p^2}{8 \pi s} \frac{\sqrt{s - 4 M_Z^2}}{\sqrt{s - 4 M_S^2}} \frac{\left(s^2 - 4 M_Z^2 s + 12 M_Z^4\right)}{\left[ \left( s - M_h^2 \right)^2 + \Gamma_h^2 M_h^2 \right]}.
 \end{equation}
 \item Annihilation into two Higgs bosons:
 \begin{multline}
  \sigma(SS \to hh) = \frac{\lambda_p^2}{16 \pi s} \frac{\sqrt{s-4 M_h^2}}{\sqrt{s- 4 M_S^2}} \left[\frac{2 \left( s + 2 M_h^2 \right)^2}{\left(s - M_h^2 \right)^2} + \frac{16 \lambda_p^2 v_0^4}{M_h^4 - 4 M_h^2 M_S^2+ M_S^2 s} \right. \\
  \left.  + \frac{32 \lambda_p v_0^2 \left( s^2 - 2 \lambda_p s v_0^2 - 4 M_h^4 + 2 \lambda_p M_h^2 v_0^2\right)}{\sqrt{s-4 M_h^2} \sqrt{s - 4 M_S^2 } \left( 2 M_h^4 - 3 M_h^2 s + s^2 \right)} \tanh^{-1}\left(\frac{\sqrt{s-4 M_h^2} \sqrt{s - 4 M_S^2 }}{2 M_h^2 - s} \right) \right].
 \end{multline}
 In the limit $s \to 4 M_S^2$, the cross section times velocity is given by
 \begin{equation}
   \sigma v_\text{rel} ( SS \to hh) = \frac{\lambda_p^2  \left[ M_h^4 - 4 M_S^4 + 2 \lambda_p v_0^2 \left( 4 M_S^2 - M_h^2 \right)\right]^2}{4 \pi M_S^2 \left( M_h^4 - 6 M_h^2 M_S^2 + 8 M_S^2 \right)^2} \sqrt{1 - \frac{M_h^2}{M_S^2}}.
 \end{equation}

\item Annihilation into $\gamma \gamma$:
 \begin{equation}
 \label{eq:generalAnnihilation}
  \sigma ( S S \to \gamma \gamma )= \frac{4 \lambda_p^2 \ v_0^2 \  \Gamma_{\gamma \gamma} (s) }{\sqrt{s-4 M_S^2}\left[\left( s - M_h^2\right)^2 + M_h^2 \Gamma_h^2 \right]},
 \end{equation}
 where the width is given by~\cite{Djouadi:2005gi}
  \begin{equation}
  \Gamma_{\gamma \gamma} (s) = \frac{\alpha^2 s^{3/2}}{256 \pi^3 v_0^2} \left| \sum_f N_c^f Q_f^2 A_{1/2}^h(\tau_f) + A_1^h (\tau_W) \right|^2,
 \end{equation}
with the form factors
\begin{align}
 A_{1/2}^h(\tau) &= 2 \left[ \tau + (\tau - 1) f(\tau) \right] \tau^{-2}, \\
 A_1^h (\tau) &= - \left[ 2 \tau^2 + 3 \tau + 3 ( 2\tau - 1) f(\tau) \right] \tau^{-2},
 \label{eq:loopFunctions}
\end{align}
and the function
\begin{equation}
\label{eq:functionf}
 f(\tau) = 
 \begin{cases}
  \text{arcsin} ^2 \sqrt{\tau} & \tau \leq 1 \\
  - \frac{1}{4} \left[ \log \frac{1 + \sqrt{1-\tau^{-1}}}{1 - \sqrt{1-\tau^{-1}}} - i\pi \right]^2 & \tau > 1
 \end{cases} .
\end{equation}
The parameters $\tau_i$ for fermions and the $W$ are given by
\begin{equation}
 \tau_f = \frac{s}{4 M_f^2} \text{ and } \tau_W = \frac{s}{4 M_W^2}.
\end{equation}
\item Annihilation into $Z \gamma$:
 \begin{equation}
  \sigma ( S S \to Z \gamma )= \frac{4 \lambda_p^2 \ v_0^2 \  \Gamma_{Z \gamma} (s) }{\sqrt{s-4 M_S^2}\left[ \left( s - M_h^2\right)^2 + M_h^2 \Gamma_h^2 \right]}\, .
 \end{equation}
The width is given by~\cite{Djouadi:2005gi}
  \begin{equation}
  \Gamma_{ Z \gamma} (s) = \frac{\alpha M_W^2 s^{3/2}}{128 \pi^4 v_0^4} \left( 1 - \frac{M_Z^2}{s}\right)^3 \left| \sum_f N_c^f \frac{Q_f \hat{v}_f}{c_W} A_{1/2}^h(\tau_f,\lambda_f) + A_1^h (\tau_W, \lambda_W) \right|^2,
 \end{equation}
with 
\begin{equation}
 \hat{v}_f = 2 I_f^3 - 4 Q_f s_W^2,
\end{equation}
and the parameters\footnote{Notice that the definition of the $\tau_i$ is the inverse of the definition used in the $h \to \gamma \gamma$ case.}
\begin{equation}
 \tau_i= \frac{4 M_i^2}{s} \text{ and } \lambda_i = \frac{4 M_i^2}{M_Z^2}.
\end{equation}
The form factors are
\begin{align}
 A_{1/2}^h (\tau, \lambda) &= \left[I_1(\tau, \lambda) - I_2 (\tau, \lambda) \right], \\
 A_1^h (\tau, \lambda) &= c_W \left\{ 4 \left( 3 - \frac{s_W^2}{c_W^2}\right) I_2 (\tau, \lambda) + \left[\left(1 + \frac{2}{\tau}\right) \frac{s_W^2}{c_W^2} - \left( 5 + \frac{2}{\tau}\right) \right] I_1(\tau, \lambda)\right\},
\end{align}
with the functions
\begin{align}
 I_1( \tau, \lambda) &= \frac{\tau \lambda}{2 ( \tau - \lambda)} + \frac{\tau^2 \lambda^2}{2 (\tau - \lambda)^2} \left[ f(\tau^{-1}) - f(\lambda^{-1})\right] + \frac{\tau^2 \lambda}{(\tau-\lambda)^2} \left[ g(\tau^{-1}) - g(\lambda^{-1})\right], \\
 I_2 (\tau, \lambda) &= - \frac{\tau \lambda}{2 (\tau - \lambda)}  \left[ f(\tau^{-1}) - f(\lambda^{-1})\right],
\end{align}
with $f(\tau)$ from Eq.~\eqref{eq:functionf} and 
\begin{equation}
 g(\tau) = 
 \begin{cases}
 \sqrt{\tau^{-1} - 1} \ \text{arcsin} \sqrt{\tau} & \tau \geq 1 \\
 \frac{\sqrt{1-\tau^{-1}}}{2} \left[ \log \frac{1+\sqrt{1-\tau^{-1}}}{1-\sqrt{1-\tau^{-1}}} - i \pi \right] & \tau < 1
 \end{cases}.
\end{equation}
\end{itemize} 
\section{Final State Radiation} 
\label{app:FSR}
In this appendix, we use 
\begin{equation}
 x_1 = \frac{E_1}{M_S} \quad \text{and} \quad x_\gamma = \frac{E_\gamma}{M_S}.
\end{equation}

\begin{itemize}
\item $SS \to \bar{f} f \gamma$: in the limit $s \to 4 M_S^2$ the the matrix element used in Eq.~\eqref{eq:FinalStateRadiation} is given by 
\begin{multline}
 \left|\mathcal{M}_\text{FSR} (SS \to \bar{f} {f} \gamma ) \right|^2= \frac{8\lambda_p^2 e^2  Q_f^2 M_f^2 N_c^f}{M_S^4 (x_1-1)^2 (x_1+x_\gamma-1)^2\left[\left(M_h^2-4M_S^2\right)^2 + \Gamma_h^2 M_h^2\right]} \\ 
 \times \Big\{M_f^4 \left(x_1-1\right)^2+M_f^2 M_S^2 \left[4 x_1^3+2 x_1^2 \left(x_\gamma-3\right)-2 x_1 \left(x_\gamma -3\right) x_\gamma + x_\gamma \left(5  x_\gamma-8\right)+2\right]  \\
   -2 M_S^4 \left(x_1-1\right) \left[\left(x_\gamma-2\right) x_\gamma+2\right]
 \left(x_1+x_\gamma-1\right) \Big\}.
\end{multline}
Integrated over $x_1$, we obtain for the differential cross section 
\begin{multline}
  \frac{d \sigma v_\text{rel}   (SS \to \bar{f} {f} \gamma ) }{dx_\gamma} =   \frac{\lambda_p^2 e^2  Q_f^2 M_f^2 N_c^f}{4 \pi^3 M_S^3 x_\gamma \left[\left(M_h^2-4M_S^2\right)^2 + \Gamma_h^2 M_h^2\right]} \\
  \times \left\{\left(M_f^2 + 2 M_S^2 \right) \sqrt{(x_\gamma -1) \left[M_f^2 + M_S^2 (x_\gamma -1 )\right]} \right. \\ 
  \left. + M_S \left[ 2 M_f^2 (x_\gamma -1) + M_S^2 \left( 2 + x_\gamma (x_\gamma -2) \right)\right] \ln \left( \frac{1 + \sqrt{1 + \frac{M_f^2}{(x_\gamma-1) M_S^2}}}{1 - \sqrt{1 + \frac{M_f^2}{(x_\gamma-1) M_S^2}}} \right)  \right\}.
\end{multline}
\item $SS \to WW \gamma$: in the limit $s \to 4 M_S^2$ the matrix element used in Eq.~\eqref{eq:FinalStateRadiation} is given by 
 \begin{multline}
 \left|\mathcal{M}_\text{FSR} (SS \to WW  \gamma) \right|^2 = \frac{4 \lambda_p^2 e^2}{M_S^4 \left( x_1 - 1 \right)^2 \left(x_1 + x_\gamma - 1 \right)^2 \left[\left(4 M_S^2 - M_h^2 \right)^2 + M_h^2 \Gamma_h^2\right]} \\
 \times \Big\{ -3 M_W^6 x_\gamma^2 + 16 M_S^6 \left( x_1 - 1 \right) \left( x_\gamma - 1 \right) \left( x_\gamma + x_1 - 1 \right)
 + 4 M_S^4 M_W^2 \left[ 4 + 4 x_1^2 \left( 1 + (x_\gamma - 1) x_\gamma \right) \right. \\
 \left. + 4 x_1 \left(x_\gamma - 2 \right) \left(1 + (x_\gamma -1) x_\gamma  \right) + x_\gamma \left( -8 + x_\gamma ( 7 +2 (x_\gamma - 2) x_\gamma )\right)\right]  \\
 + 4 M_S^2 M_W^4 \left[ -3 + 3 x_1^2 \left( x_\gamma - 1 \right) + 3 x_1 \left( x_\gamma - 2 \right) \left( x_\gamma - 1 \right) - 2 \left( x_\gamma -3 \right) x_\gamma \right]\Big\}.
 \end{multline}
Integrated over $x_1$, we obtain for the differential cross section 
\begin{multline}
  \frac{d \sigma v_\text{rel}  (SS \to W W \gamma )}{dx_\gamma}= \frac{e^2 \lambda_p^2 }{8 \pi ^3 M_S^4 x_\gamma
  \left[\left(4 M_S^2 - M_h^2 \right)^2 + M_h^2 \Gamma_h^2\right]}\\
  \times \Bigg\{ 2 M_S \sqrt{(x_\gamma -1 ) \left[ M_W^2 + M_S^2 (x_\gamma -1) \right]} \left[ 4 M_S^2 M_W^2 - 3 M_W^4 + 4 M_S^4 \left(2 x_\gamma^2 -1 \right) \right]  \\
  \left. - \left( 4 M_S^4 - 4 M_S^2 M_W^2 + 3 M_W^4 \right) \left[ M_W^2 + 2 M_S^2 (x_\gamma -1) \right] \ln \left( \frac{1 + \sqrt{1 + \frac{M_W^2}{(x_\gamma-1) M_S^2}}}{1 - \sqrt{1 + \frac{M_W^2}{(x_\gamma-1) M_S^2}}} \right) \right\} .
\end{multline}
\end{itemize}
%
\section{Gamma-Ray Spectrum from Dark Matter Annihilation}
\label{app:gammaspectrum}
%
We derive the expression for the differential cross section of dark matter annihilation to a photon and an unstable particle $X$. 
Consider the annihilation $S S \rightarrow X + \gamma \rightarrow f \bar{f} \gamma$. 
We begin by decomposing the three-body phase space of the final states into two two-body phase space parts and an integral over the mediator as
\begin{align}
& d\Pi(\sqrt{s}; p_1,p_2, p_\gamma)   =  \frac{d^3 p_1}{(2\pi)^3 2 E_1}\frac{d^3 p_2}{(2\pi)^3 2 E_2}\frac{d^3 p_\gamma}{(2\pi)^3 2 E_\gamma} (2\pi)^4 \delta(k_1 + k_2 -p_1 - p_2 - p_\gamma) \nonumber \\
& =  \frac{d^4 p}{(2\pi)^4} \frac{d^3 p_\gamma}{(2\pi)^3 2 E_\gamma} (2\pi)^4 \delta(k_1 +k_2  -p -p_\gamma) \frac{d^3 p_1}{(2\pi)^3 2 E_1}\frac{d^3 p_2}{(2\pi)^3 2 E_2} (2\pi)^4 \delta(p -p_1 -p_2), 
\end{align}
where $p_1$ and $p_2$ are the momenta of $f$ and $\bar{f}$, $p_\gamma$ is the photon momentum, $p$ the off-shell momentum of $X$ and $M^2$ 
the invariant mass of $p_1$ and $p_2$ combined. $k_1$ and $k_2$ are the momenta of the incoming particles.
Using the relations  
$\frac{\partial M^2}{\partial |p|}= \frac{\partial }{\partial |p|}(p_0^2- |p|^2 ) = -2|p| $ and $E_p dE_p = |p| d|p| $ one finds that $dM^2 = -2 E_p  dE_p$ and
\begin{align}
d\Pi(\sqrt{s}; p_1,p_2, p_\gamma) & =  - \frac{dM^2}{2\pi} \frac{d^3 p}{(2\pi)^3 2 E_p} \frac{d^3 p_\gamma}{(2\pi)^3 2 E_\gamma} (2\pi)^4 \delta(k_1 +k_2  -p -p_\gamma)  \nonumber \\ 
& \quad \times \frac{d^3 p_1}{(2\pi)^3 2 E_1}\frac{d^3 p_2}{(2\pi)^3 2 E_2} (2\pi)^4 \delta(p -p_1 -p_2)  \nonumber \\
& =  - \frac{d M^2}{2 \pi} \,  d\Pi(\sqrt{s}; p , p_\gamma) \, d\Pi(p ; p_1,p_2).
\end{align}
With this decomposition, the cross section can be written as
\begin{eqnarray}
\label{eq:diffCrossSectionMass}
 d \sigma  & = &  \frac{|\mathcal{M}|^2}{\Phi_{S S}}  d\Pi(\sqrt{s}, p_1,p_2, p_\gamma)  = - \frac{|\mathcal{M}|^2}{\Phi_{S S}} \frac{d M^2}{2 \pi} \,  d\Pi(\sqrt{s}; p , p_\gamma) \, d\Pi(p ; p_1,p_2),
\end{eqnarray}
where $\Phi_{S S}$ is the dark matter flux.
Now, we write the differential amplitude of the process assuming a narrow width of the intermediate particle.
Then, one finds  
\begin{eqnarray} 
 \frac{d \sigma}{d M^2}  & =&  - \frac{1}{2 \pi} \left| \frac{1}{M^2-m_X^2 + i \Gamma_X m_X}\right|^2 \int_{p , p_\gamma} \frac{|\mathcal{M}_{S S \rightarrow X \gamma}|^2}{\Phi_{S S}}  \, d\Pi(\sqrt{s}; p, p_\gamma)  \int_{p_1, p_2} |\mathcal{M}_{X\rightarrow q  \bar{q}}|^2  d\Pi(p; p_1, p_2) \nonumber \\ 
& = & -\frac{\sigma (S S \rightarrow X \gamma) \, \Gamma_X m_X }{\pi} \left| \frac{1}{M^2-m_X^2 + i \Gamma_X m_X}\right|^2 \,.
\end{eqnarray}
Using the fact that $M^2 = 4 M_S^2 - 4 M_S E_\gamma $, which gives a Jacobian factor of $\frac{dM^2}{dE_\gamma} = - 4 M_S$, we obtain the final result for the differential cross section
\begin{align}
\label{eq:DiffCrossSectionUnstableParticle}
\frac{d \sigma}{d E_\gamma} = \frac{\sigma (S S \rightarrow X \gamma)}{\pi}\frac{4  M_S m_X \Gamma_X}{(4M_S^2 - 4 M_S E_\gamma -m_X^2)^2 + \Gamma_X^2 m_X^2} \equiv \sigma (S S \rightarrow X \gamma) \frac{dN_\gamma}{dE_\gamma}\,.
\end{align}
%


\end{document}